\documentclass[10pt,conference]{IEEEtran}
\IEEEoverridecommandlockouts

\usepackage[T1]{fontenc}
\usepackage[utf8x]{inputenc}
\usepackage{amsmath,amssymb,amsfonts}
\usepackage{mathtools}
\usepackage{graphicx}
\usepackage{booktabs}
\usepackage{xspace}
\usepackage{multirow}
\usepackage{xcolor}
\usepackage[colorlinks=true,urlcolor=teal,
            linkcolor=teal,citecolor=teal]{hyperref}\usepackage{cleveref}
\usepackage{balance}
\usepackage{enumitem}
\usepackage{url}
\usepackage{csquotes}
\usepackage{glossaries-extra}
\usepackage{tikz}
\usepackage{orcidlink}
\usetikzlibrary{patterns,decorations.pathmorphing,calc,positioning}
\setabbreviationstyle[acronym]{long-short}
\glssetcategoryattribute{acronym}{nohyperfirst}{true}

\usepackage[backend=biber,style=ieee,maxnames=3,maxcitenames=2, minbibnames=3,mincitenames=1,maxbibnames=3,citetracker=true,citestyle=ieee-comp,defernumbers=true]{biblatex}

\newbibmacro*{title+link}[1]{%
  \iffieldundef{doi}
    {%
      \iffieldundef{url}
        {#1}%
        {\href{\thefield{url}}{\textcolor{teal}{#1}}}%
    }
    {\href{https://doi.org/\thefield{doi}}%
       {\textcolor{teal}{#1}}}%
}

\DeclareFieldFormat
  [online,article,inbook,incollection,inproceedings,patent,thesis,unpublished,misc,book]
  {title}{%
    \usebibmacro{title+link}{#1}%
  }

\renewbibmacro*{doi+eprint+url}{}
\renewbibmacro*{url+urldate}{}

\DeclareFieldFormat{pages}{}
\AtEveryBibitem{
	\clearfield{pages}
	\clearfield{isbn}
	\clearfield{eprint}
	\clearfield{issn}
	\clearfield{abstract}
	\clearfield{volume}
	\clearfield{number}
	\clearfield{month}
  \clearfield{day}
  \clearfield{language}
  \clearfield{note}
}

\renewbibmacro{in:}{}
\AtBeginBibliography{\footnotesize}

\newacronym{ftqc}{FTQC}{Fault Tolerant Quantum Computers}
\newacronym{icc}{ICC}{Intraclass Correlation Coefficient}
\newacronym{nisq}{NISQ}{Noisy Intermediate Scale Quantum}
\newacronym{pec}{PEC}{Probabilistic Error Cancellation}
\newacronym{qec}{QEC}{Quantum Error Correction}
\newacronym{qem}{QEM}{Quantum Error Mitigation}
\newacronym{qft}{QFT}{Quantum Fourier Transformation}
\newacronym{qtc}{QTC}{Quantum Trotter Circuits}
\newacronym{zne}{ZNE}{Zero-Noise Extrapolation}

\usepackage{censor}
\newboolean{anonymous}
\setboolean{anonymous}{false}
\ifbool{anonymous}{}{}
\ifbool{anonymous}{}{\renewcommand{\blackout}[1]{#1}}
\ifbool{anonymous}{\newcommand{\genemail}[2]{\href{mailto:xxx.xxx@xx.xx}{\blackout{#2}}}}{\newcommand{\genemail}[2]{\href{#1}{#2}}}

\newcommand{\repro}{\href{https://github.com/lfd/qce26_claim_against_measurement}{reproduction package\xspace}}
\newcommand{\zenodo}{\href{https://doi.org/10.5281/zenodo.21538274}{Zenodo archive\xspace}}

\newcommand{\na}{\textcolor{gray}{n/a}}

\newcommand{\etal}{\emph{et al.}\xspace}
\newcommand{\eg}{\emph{e.g.}\xspace}

\begin{document}

\bstctlcite{BSTcontrol}

\title{Claim against Measurement: Statistical Artefacts\\ in Quantum Error Mitigation Benchmarks}

\author{
    \IEEEauthorblockN{\blackout{Dominik Köster\orcidlink{0009-0001-0230-3123}}}
    \IEEEauthorblockA{
        \blackout{\textit{Technical University of}} \\
        \blackout{\textit{Applied Science Regensburg}} \\
        \blackout{Regensburg, Germany} \\
        \genemail{mailto:dominik.koester@othr.de}{dominik.koester@othr.de}
    }
    \and
    \IEEEauthorblockN{\blackout{Wolfgang Mauerer\orcidlink{0000-0002-9765-8313}}}
    \IEEEauthorblockA{
        \blackout{\textit{Technical University of}} \\
        \blackout{\textit{Applied Science Regensburg}} \\
        \blackout{\textit{Siemens AG, Foundational Technology}} \\
        \blackout{Regensburg/Munich, Germany} \\
        \genemail{mailto:wolfgang.mauerer@othr.de}{wolfgang.mauerer@othr.de}
    }
}

\maketitle
\begin{abstract}
\gls{qem} is widely regarded as a plausible bridge from \gls{nisq}
devices to \gls{ftqc}. Yet the empirical studies used to assess the
effectiveness of \gls{qem} techniques on concrete problems
have received comparatively little scrutiny with respect to the validity
of their conclusions. We systematically review 81 recent \gls{qem}
papers using an eight-criterion framework covering statistical rigour,
reproducibility, and reporting quality. Among the 59 papers for which
statistical evidence is applicable, only 15 (25\%) use inferential
methods, while 25 (42\%) report uncertainty only descriptively, without
testing whether the claimed effects are statistically supported.

To demonstrate the consequences of these omissions, we use
\gls{zne} as a representative and widely used case study and identify
two compounding sources of artefacts in current \gls{qem} benchmarks.
First, we observe \emph{parameter sensitivity}: in a 132-configuration
sweep, implicitly assumed choices such as scale factors, extrapolation
method, and hardware calibration are not merely incidental but
\emph{active}, with variations changing conclusions from statistically
significant improvement to statistically significant degradation.
Second, we identify a \emph{drift-induced effectiveness illusion}: in a
72-hour longitudinal study on real hardware, temporal drift alone
can make the same \gls{zne} configuration exhibit an effect size more than three times as large, depending solely on when it is executed, and also 
drastically reduces the effective number of independent observations. These findings do not imply that \gls{qem} methods are intrinsically unsound; rather, they show that current evaluation
practice can make mitigation performance appear more robust than the
evidence warrants. We therefore propose minimum reporting standards
for \gls{qem} evaluations, including explicit parameter documentation,
robustness checks, longitudinal drift assessment, and inferential
statistical testing with effect-size reporting.
\end{abstract}


\section{Introduction}
\label{sec:introduction}
We remain in the \gls{nisq} era of quantum computing~\cite{preskill_quantum_2018,felix:23:imperfections,simon:24:noise,bharti:2022}, still some distance from \gls{ftqc} architectures with sufficiently many qubits, low error rates, and operational \gls{qec}~\cite{preskill_fault-tolerant_1997,preskill_beyond_2025}. Although \gls{qec} is clearly the long-term route to reliable quantum computation, present implementations still incur substantial qubit overheads~\cite{beverland_surface_2022,gidney_yoked_2023}. In this intermediate regime, \gls{qem} has emerged (among hardware-efficient problem formulations~\cite{lukas:25:sat,lukas:24:reduction,schmidbauer:2026} or target design automation techniques~\cite{simon:25:designautomation,wille:2024}) as a pragmatic approach for reducing the impact of noise~\cite{stefan:25:noise} on current \gls{nisq} devices without implementing full error correction~\cite{temme_error_2017,li_efficient_2017,endo_practical_2018,cai_quantum_2023}. Techniques such as \gls{zne}~\cite{temme_error_2017,li_efficient_2017}, \gls{pec}~\cite{temme_error_2017,berg_probabilistic_2023}, and Clifford data regression~\cite{czarnik_error_2021} have shown promise in improving the accuracy of quantum computations. Beyond dedicated benchmarking studies, \gls{qem} techniques are increasingly adopted as standard components in broader quantum computing research~--~from variational quantum eigensolvers and quantum structure simulations~\cite{kandala_error_2019,dumitrescu2018cloud} via optimisation~\cite{lukas:26:quadratisation,Lucas2014,tom:25:loop,lukas:25:path,schoenberger:2023} and machine learning~\cite{schuld2015introduction,franz:2024,schuld2021machine,wittek2014quantum} to quantum simulation of many-body systems~\cite{chen2022error} and demonstrations of evidence for quantum utility~\cite{kim_evidence_2023}~--~where framework-default parameters are silently inherited and their influence is not examined. Artefacts arising from such implicitly assumed parameter choices therefore propagate beyond \gls{qem} evaluation into any higher-level result that relies on mitigation as a building block. However, the empirical evaluation of \gls{qem} techniques themselves frequently rests on experiments whose statistical foundations are not always commensurate with the strength of the conclusions drawn from them.

Previous work has already identified important statistical challenges in the evaluation of \gls{qem} techniques, including the need for explicit hypothesis testing~\cite{saki_hypothesis_2023}. In this work, however, our systematic review of 81 recent \gls{qem} papers (Section~\ref{sec:review}) shows that such concerns remain largely unaddressed in practice: most papers rely on descriptive reporting rather than inferential statistical evidence. This is not merely a matter of presentation, but one with direct experimental consequences. Reported \gls{qem} improvements are often small, since current \gls{nisq} hardware constrains experiments to shallow circuits and shot budgets limited by cost, placing many studies in regimes where the true effect is modest at best~\cite{russo_testing_2023,maupin_error_2024}. In such settings, shot noise, hardware drift, and implicitly assumed parameter choices can each be sufficient to alter the experimental conclusion. Without careful inferential testing, awareness of influence factors and boundary conditions, and the artefacts needed for faithful reproduction, neither the original study nor a replication can reliably separate genuine mitigation effects from artefacts.

In this paper, we identify two compounding sources of artefacts in \gls{qem} evaluation, each independently capable of producing misleading results. First, a systematic replication study shows that documented parameters cover only a small portion of an experiment's parameter space: unspecified choices~--~such as scale factors, folding strategy, or calibration snapshot~--~are \emph{active}, meaning their variation can shift the outcome from significant improvement to significant worsening relative to the unmitigated baseline. Second, a longitudinal hardware study reveals that temporal drift alone can produce large variation in apparent \gls{zne} effectiveness on the same device at different access times. Given the dominant use of cloud services with indeterministic batch access for many quantum experiments, this variability poses a severe challenge for both reproducing and interpreting empirical results. We focus on \gls{zne} with Richardson extrapolation as a representative case, as it is the most widely reported \gls{qem} technique in our corpus, and propose a minimum reporting checklist for \gls{qem} evaluations.

Concretely, our contributions are as follows:
\begin{itemize}
  \item We provide a systematic eight-criterion review of 81 \gls{qem} papers, revealing that the majority of papers lack inferential statistics and drift control.
  \item We introduce a replication pipeline and case study demonstrating that implicitly assumed \gls{zne} parameters are \emph{active}: their variation shifts experimental outcomes across a large fraction of tested configurations.
  \item A longitudinal drift study shows that temporal hardware drift produces a \emph{drift-induced effectiveness illusion}, where apparent \gls{zne} effectiveness varies substantially across identical experiments at different times.
  \item We derive minimum reporting standards for \gls{qem} benchmarks to avoid incorrect claims and misinterpretations of \gls{qem} experiments.
\end{itemize}

\section{Related Work}
\label{sec:related}
Cai~\etal~\cite{cai_quantum_2023} provide a comprehensive review of \gls{qem} techniques, including their theoretical foundations, practical implementations, and performance analysis. Takagi~\etal~\cite{takagi_fundamental_2022} derive fundamental sampling-overhead bounds under global depolarising noise, generalised by Quek~\etal~\cite{quek_exponentially_2024} for local noise. Krebsbacher~\etal~\cite{krebsbach_optimization_2022} derive variance bounds for Richardson extrapolation, guiding scale factor selection to minimise amplification. On the statistical side, Saki~\etal~\cite{saki_hypothesis_2023} propose a hypothesis-testing framework for \gls{qem} evaluation, while Li~\etal~\cite{li_methodological_2026} survey the use of statistical methods in quantum software testing, finding that formal statistical methods are similarly rare in that domain. Moguel~\etal~\cite{moguel_quantum_2024} review quantum benchmarking methods, proposing a quantum experiment guideline extending established best practices from classical software benchmarking. Garmon~\etal~\cite{Garmon_2020} propose treating error mitigability itself as a first-class device benchmark~--~a view our results support. Our empirical case study targets \gls{zne} with Richardson extrapolation, the most prevalent technique in our corpus; structurally different methods such as \gls{pec}, whose sampling overhead grows with noise~\cite{berg_probabilistic_2023}, Clifford data regression~\cite{czarnik_error_2021}, or learning-based mitigation may exhibit different parameter-sensitivity and drift profiles, so our quantitative magnitudes need not transfer directly~--~though the underlying reporting gaps (C3, C4) remain method-agnostic. 
On reproducibility, Senapati~\etal~\cite{senapati_towards_2023, senapati_pqml_2024} highlight the reproducibility challenges in quantum machine learning, being device variability and temporal drift. Hirasaki~\etal~\cite{hirasaki_detection_2023} demonstrate temporal fluctuations in superconducting qubits, yielding different measurements over time points. 

\section{Background}
\label{sec:background}

\subsection{Zero-Noise Extrapolation (ZNE)}
\gls{zne}, first introduced by Li and Benjamin~\cite{li_efficient_2017} and Temme~\etal~\cite{temme_error_2017}, is a widely used \gls{qem} technique that estimates the noise-free result of a quantum computation by extrapolating results obtained at amplified noise levels. Given the noise scale $\lambda$, the result of the smallest error rate on a circuit is given by the expectation value $E(\lambda_1)$ \cite{cai_quantum_2023, majumdar_best_2023}. By artificially increasing the noise to levels $\lambda_1 < \lambda_2 < \dots < \lambda_K$~--~called \emph{scale factors}~--~we can fit a model to extrapolate the noise-free expectation value $E(0)$. Common amplification strategies include pulse stretching~\cite{temme_error_2017, kandala_error_2019}, unitary folding~\cite{giurgica-tiron_digital_2020}, and gate-level folding~\cite{hour_improving_2024}. To compute the extrapolated value, besides standard approaches like linear~\cite{li_efficient_2017}, polynomial and exponential extrapolation~\cite{endo_practical_2018, majumdar_best_2023}, \emph{Richardson extrapolation} is a commonly employed method \cite{temme_error_2017,endo_practical_2018, giurgica-tiron_digital_2020, krebsbach_optimization_2022} that fits a polynomial of degree $K-1$ through $K$ data points. The zero-noise limit is in this case estimated~\cite{cai_quantum_2023} as \(\hat{E}(0) = \sum_{k=1}^K c_k E(\lambda_k)\), where \(\sum_{k=1}^K c_k = 1\), and coefficients $c_k$ are determined by Lagrange interpolation.

Since $\text{Var}(\hat{E}_\text{ZNE}) = \sum_{k=1}^K c_k^2 \, \text{Var}(\hat{E}(\lambda_k))$, a convenient pre-experiment bound on variance amplification is $\sum_{k=1}^{K}|c_k|$~\cite{cai_quantum_2023, krebsbach_optimization_2022}~--~computable from the scale factors alone, without running any circuits. For the widely used default set $\{1, 3, 5\}$~\cite{giurgica-tiron_digital_2020, majumdar_best_2023}, $\sum_{k=1}^K|c_k| = 3.5$, whereas $\{1, 1.1, 1.25, 1.5\}$~--~motivated by arguments that finer spacing reduces extrapolation error~\cite{krebsbach_optimization_2022}~--~yields $\sum_{k=1}^K|c_k| = 681$: a $194\times$ difference in the variance bound. Scale factor choice alone can therefore dominate the variance budget of a \gls{zne} experiment~--~motivating its treatment as an active parameter in Section~\ref{sec:experiments_khan}.

\subsection{Statistical Methods}
\label{sec:bg-stats}
Our approach, implemented in a \href{https://github.com/lfd/qce26_claim_against_measurement}{reproducible pipeline} (link in PDF)~\cite{mauerer_repro_2022}, relies on standard frequentist tools to quantify how much a \gls{qem} technique improves performance. Unlike clinical trials or psychology, \gls{qem} evaluation has no domain-specific statistical standards: no agreed effect-size threshold for claiming practical mitigation, no canonical hypothesis test, and no prescribed power requirement. We therefore adopt well-validated general-purpose tools from empirical science~\cite{cohen_statistical_1988, fahrmeir_statistik_2023, devore2008probability}~--~standard there, yet still rare in quantum computing. As they are textbook knowledge, we only briefly review their essential properties in the \gls{qem} context.

\paragraph{Paired $t$-test}
A paired $t$-test compares differences between two groups: Given $n$ independent repetitions producing raw errors $|\epsilon_{\text{raw},i}|$ and mitigated errors $|\epsilon_{\text{mit},i}|$, the paired difference $\delta_i = |\epsilon_{\text{raw},i}| - |\epsilon_{\text{mit},i}|$ captures per-repetition improvement.
The paired $t$-test evaluates the null hypothesis $H_0\!: \mu_\delta = 0$ via $t = \bar{\delta} / (s_\delta / \sqrt{n})$. We classify each configuration as \emph{significantly better} ($p < 0.05$, $d > 0$), \emph{not significant} ($p \geq 0.05$), or \emph{significantly worse} ($p < 0.05$, $d < 0$) based on a 5\% significance level (while using a prescribed explicit significance level is known to be exhibit a number of issues~\cite{wasserstein2016asa}, we stick to this approach as it is common practice in the considered references, and conclusion stability can only be assessed when the same approach as in the original work is employed). As a non-parametric alternative that does not assume normality of the paired differences, we compute the Wilcoxon signed-rank test for every configuration.

\paragraph{Effect-Size Measures}
We use two complementary effect-size quantities: \textbf{Cohen's $d$} $= \bar{\delta} / s_\delta$~\cite{cohen_statistical_1988} is the standard effect-size measure in empirical sciences such as clinical trials or psychology, with well-established conventions $|d| = 0.2/0.5/0.8$ (small/medium/large)~\cite{cohen_statistical_1988} and $1.2/2.0$ (very large/huge)~\cite{sawilowsky_new_2009}; positive $d$ means \gls{qem} reduces, negative $d$ means it increases error. These thresholds were, however, calibrated for domains where effect sizes are bounded by natural variability rather than a controllable experimental parameter. In shot-based quantum experiments, $s_\delta \propto 1/\sqrt{n_\text{shots}}$, so $d$ scales with shot count: at $n_\text{shots} = 4096$, a modest improvement of $\Delta = 0.18$ already yields $d \approx 6$, far above the established thresholds. Absolute $d$ values are not comparable to those conventions and we use $d$ primarily as a relative metric for cross-configuration comparison. Because no single effect-size measure is universally agreed on for \gls{qem} evaluation, we additionally report \textbf{Cliff's $\delta$} $= (N_{>} - N_{<}) / n$~--~where $N_{>}$ and $N_{<}$ count paired differences $\delta_i > 0$ and $\delta_i < 0$~--~as a non-parametric, distribution-free alternative in $[-1, +1]$ that is insensitive to this shot-count inflation and directly reflects the probability of a genuine directional improvement across repetitions. Intuitively, Cohen's $d$ expresses improvement in units of its own scatter, whereas Cliff's $\delta$ counts how often mitigation helps rather than hurts across repetitions.

\section{Systematic Review}\label{sec:review}
To evaluate the current state of statistical reporting in \gls{qem} related papers and to identify common pitfalls and areas for improvement, we systematically reviewed $81$ publications published over half a decade (2022--2026), collected via Google Scholar, arXiv, IEEE digital libraries, and forward/backward citation tracking from the review by Cai~\etal~\cite{cai_quantum_2023}. We used queries like \enquote{quantum error mitigation}, \enquote{zero-noise extrapolation}, \enquote{probabilistic error cancellation} in combination with terms like \enquote{algorithm}, \enquote{drift}, or \enquote{experiment} to identify relevant papers (the full list is provided in the \repro{}). Each paper was evaluated based on eight criteria detailed in Table~\ref{tab:criteria} that assess the presence and quality of statistical reporting in the experimental evaluation of \gls{qem} techniques. Each criterion is rated as \emph{adequate} (clear, specific evidence), \emph{partial} (mentioned but incomplete), \emph{missing}, or \emph{not applicable}. The criteria themselves are derived from a priori expert knowledge, typical desiderata in the literature, and requirements documented in reviews~\cite{murillo_roadmap_2025,carbonelli_challenges_2024,yue_challenges_2023} or existing work on quantum reproducibility~\cite{veiga_repro_2025,gierisch_qef_2026,mauerer_repro_2022}.

\begin{table}[htbp]
  \centering
  \caption{Eight-criterion analysis framework. Each paper is rated adequate, partial, missing, or \na{}.}
  \label{tab:criteria}
  \small
  \begin{tabular}{@{}clp{5.1cm}@{}}
    \toprule
    \textbf{ID} & \textbf{Criterion} & \textbf{Key Question(s)} \\
    \midrule 
    C1 & Sample Size & How many shots/circuits? Size justified? \\
    C2 & Variance & Error bars, CIs, or variance reported? \\
    C3 & Stat.\ Evidence & Inferential or descriptive evidence for claims? \\
    C4 & Drift Control & Temporal hardware drift accounted for? \\
    C5 & Overhead & Classical/quantum overhead quantified? \\
    C6 & Noise Model & Noise model validated or discussed? \\
    C7 & Reproducibility & Code, data, or sufficient detail? \\
    C8 & Neg.\ Results & Failure cases or limitations reported? \\
    \bottomrule
  \end{tabular}
\vspace{-1em}
\end{table}

\paragraph{Review process}
The process follows established patterns for lightweight semi-formal reviews~\cite{kitchenham2007guidelines}, and aims at balancing required manual human effort and comprehensiveness/completeness of coverage.
Papers were included if they present an empirical evaluation of a \gls{qem} technique and were selected in the order returned by these libraries, which rank results by their internal relevance metrics (typically driven by citation and access counts). This sample is therefore not an exhaustive census of all \gls{qem} papers published in this period, but provides a broad, relevance-weighted cross-section. Two raters (the authors) jointly reviewed all 81 papers against the criteria, and two additional raters each independently rated a subset of 15 papers. Before consensus, raters agreed on approximately $83\%$ of paper--criterion pairs; the remaining discrepancies were resolved by consensus discussion.

Additionally, we established a comparison baseline using 
automatic text processing: A regular expression scanner matched textual evidence against patterns targeting key statistical terms~--~for example, $p$-value reporting phrases, mentions of confidence intervals, or software repository URLs~--~and an LLM (Claude Opus 4.6) filtered candidate ratings for known false positives such as physical noise probabilities misidentified as statistical $p$-values, or framework citations misidentified as reproduction packages. Automated ratings agreed with the human consensus in $77\%$ of applicable paper-criterion pairs. This automated pass serves only as a coverage cross-check and does not feed into any reported result; the human consensus is the sole basis for our analysis. Full scan logs, rating sets, and a per-paper LLM evidence report are provided in the \repro. Overall, we believe the approach provides a sound and sufficiently large-scale basis that leads to generalisable conclusions.

\begin{figure}[htbp]
  \includegraphics{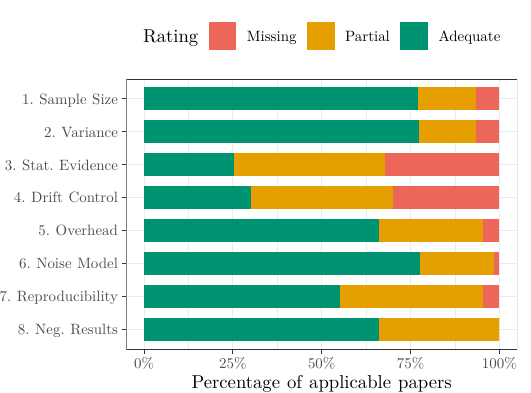}
  \caption{Summary of the systematic review results across the eight criteria.}
  \label{fig:criteria}
\end{figure}

Figure \ref{fig:criteria} shows the observed compliance of the $81$ papers with each of the eight criteria. Five criteria are well addressed, with over $60\%$ adequate reporting: Sample Size (C1, $77\%$), Variance (C2, $77\%$), Noise Model (C6, $78\%$), Overhead (C5, $66\%$), and Negative Results (C8, $66\%$). Reproducibility (C7, $55\%$) is in the middle, but plays a crucial role as it is often the only way to confirm and build upon reported results. The two criteria with the lowest compliance are Drift Control (C4, $30\%$) and Statistical Evidence (C3, $25\%$).

Of the 59 applicable papers for statistical evidence, only 15 employ inferential statistical methods: hypothesis tests, Bayesian inference, bootstrap-based comparisons, or scaling analysis with uncertainty quantification. The remaining $42\%$ report uncertainty only descriptively (error bars, standard deviations, improvement factors), and 19 papers ($32\%$) do not provide statistical evidence.

The \enquote{partial} rating on C3 deserves explanation and credit for the respective studies. These papers report variance, quantitative improvement metrics, accuracy thresholds, or bootstrap error bars, all forms of descriptive statistical evidence common in quantum experiments, but do not use them for inferential comparison.

\gls{qem} improvements are often small \cite{russo_testing_2023,maupin_error_2024}: in regimes where noise, drift, and parameter choice can each be sufficient to shift the outcome of an experiment, simple descriptive evidence alone is insufficient to confirm genuine effects. The two empirical analyses below demonstrate what practical consequences arise from a failure~--~as is omnipresent in the literature~--~to provide more complete descriptions and a more rigorous statistical analysis. Both artefact sources are independently capable of producing misleading evaluations, and the two factors compound when present simultaneously.

\section{The Reproduction Parameter Space \& Extraction Pipeline}
\label{sec:repro}

\begin{figure}[htbp]
  \centering
  \includegraphics[width=0.72\linewidth]{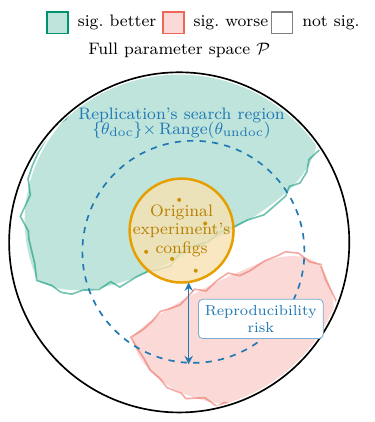}
  \caption{The full parameter space $\mathcal{P}$ of a \gls{qem} experiment, with regions of improvement (green), no improvement (grey), and worsening (red). Studies typically test a small subset of parameter configurations (orange).}
  \label{fig:param-space}
\end{figure}

Our analysis shows C3 and C4 have the smallest compliance, yet matter most when parameter choices and timing can influence or even determine the outcome of an experiment. Before we demonstrate this experimentally, let us formalise the structure of the problem. Figure~\ref{fig:param-space} illustrates the relation between a quantum (software) experiment and an attempt to reproduce or replicate the findings. Given the overall variability and currently fast-paced change in quantum hardware, rapidly changing software environments~\cite{carbonelli_challenges_2024,gierisch_qef_2026,qiskit2024}, and other factors, an exact reproduction is rarely possible even if the original software artefacts are available, which leads to an effectively different set of empirical parameters. The space of relevant parameters for quantum experiments can be divided into three categories of outcomes for a given configuration: regions of significant improvement, regions of significant worsening, and regions with no statistically significant change. In typical experiments, only a specific subset of possible values is tested per parameter (highlighted as the orange circle), where some are explicitly specified ($\theta_\text{doc}$) and others are left unspecified ($\theta_\text{undoc}$). A parameter is \emph{inert} if varying it does not change the experimental conclusion, and \emph{active} if it does. 

\subsection{Parameter Space $\mathcal{P}$}
A \gls{qem} experiment requires many choices beyond the circuit and observable.
We formalise the \emph{reproduction parameter space} for a \gls{zne} experiment as:
\( \mathcal{P} = \mathcal{H} \times \mathcal{C} \times \mathcal{Q} \times \mathcal{F} \times \mathcal{E} \times \mathcal{S}\).
As $|\mathcal{P}| > 10^{4}$, the size of the parameter space for reproductions can easily grow to hundreds of reasonable configurations.

\begin{table}[htbp]
  \centering
  \caption{Reproduction parameter space.}
  \label{tab:param-space}
  \small
  \begin{tabular}{@{}clp{4.6cm}@{}}
    \toprule
    \textbf{Axis} & \textbf{Name} & \textbf{Examples} \\
    \midrule
    $\mathcal{H}$ & Hardware/Backend & QPU vendor/model, noise model, calibration snapshot \\
    $\mathcal{C}$ & Circuit & Qubit count, depth, transp.~level \\
    $\mathcal{Q}$ & Shots \& reps & Shot count, number of repetitions \\
    $\mathcal{F}$ & Folding & Local-left, local-right, global \\
    $\mathcal{E}$ & Extrapolation & Linear, polynomial, Richardson, $\ldots$\\
    $\mathcal{S}$ & Scale factors & $\{1,3,5\}$, $\{1,2,3\}$, $\{1,1.5,\ldots,3\}$ \\
    \bottomrule
  \end{tabular}
\end{table}

\subsection{Reproduction Pipeline}
\label{subsec:pipeline}
Following established terminology, we distinguish \emph{reproduction}, where a different team re-runs the same experimental artefacts from \emph{replication}, where a different team uses a different experimental setup, yet addresses the same question as an existing piece of research. Our case study addresses both aspects: we use the original circuit specification but systematically vary the unspecified parameters across a wider range than explored in the original study.
To systematically explore the parameter space, we apply a four-stage pipeline to a target paper: (1) \textbf{Parameter Extraction}: extract all documented parameters $\theta_\text{doc}$ and identify unspecified ones $\theta_\text{undoc}$. Also categorise any parameters $\theta_\text{amb}$ that are either ambiguously mentioned or very likely given the context, but not explicitly stated. (2) \textbf{Parameter-Space Sampling}: sample $\theta_\text{undoc}$ and $\theta_\text{amb}$ across reasonable values. Reasonable is context-dependent, but in most cases, refers to the most commonly used values in literature. The result is a set of configurations $\{p_1, p_2, \ldots, p_n\} \subset \mathcal{P}$. (3) \textbf{Statistical Analysis}: run $n_\text{reps}$ repetitions for each configuration and perform statistical tests e.g. conduct paired $t$-tests, and compute Cohen's $d$ effect size. (4) \textbf{Artefact Classification}: Classify each configuration based on the statistical analysis. If the claimed improvement only holds for a subset of $\mathcal{P}$, it is not generalisable and relies on \emph{active} parameters. This pipeline is designed to be extensible to any \gls{qem} by replacing the \gls{zne} specific parameters ($\mathcal{F}, \mathcal{E}, \mathcal{S}$).

\section{The Parameter Space is Active}
\label{sec:experiments_khan}
To test whether implicitly assumed parameters $\theta_\text{undoc}$ and the statistical test gap (C3) have practical consequences~--~that is, whether an experiment with seemingly the same prerequisites can yield different outcomes~--~we demonstrate how varying different axes of $\mathcal{P}$ can shift the experimental outcome. 

\subsection{Case Study: Khan~\etal (2024)}
\subsubsection{Parameter Extraction}
To evaluate the effectiveness of different \gls{qem} techniques for \gls{nisq} devices, Khan~\etal~\cite{khan_error_2024} apply dynamic decoupling, twirled readout error extraction and \gls{zne} to four-qubit \gls{qtc} whose gate structure (\cite{khan_error_2024}, Algorithm 1) corresponds to the Trotterisation of a transverse-field Ising chain ($\mathcal{C}$). The paper executes these circuits on IBM Kyoto and Osaka machines (both now retired~\cite{ibm_retired}), and compares them with an ideal QASM Simulation ($\mathcal{H}$)~\cite{cross_openqasm_2022}. Error rates, qubits properties and architecture (at the time of execution) are documented, as well as pseudo-code for the \gls{qem} procedures. Unfortunately, parameters are neither explicitly nor implicitly (via a reproduction package) available. This leaves us with the unknown  $\mathcal{F}$, $\mathcal{E}$, $\mathcal{S}$, and $\mathcal{Q}$.

\subsubsection{Parameter-Space Sampling} \label{subsec:khan-params}

\begin{figure*}[htbp]
  \centering
  \includegraphics{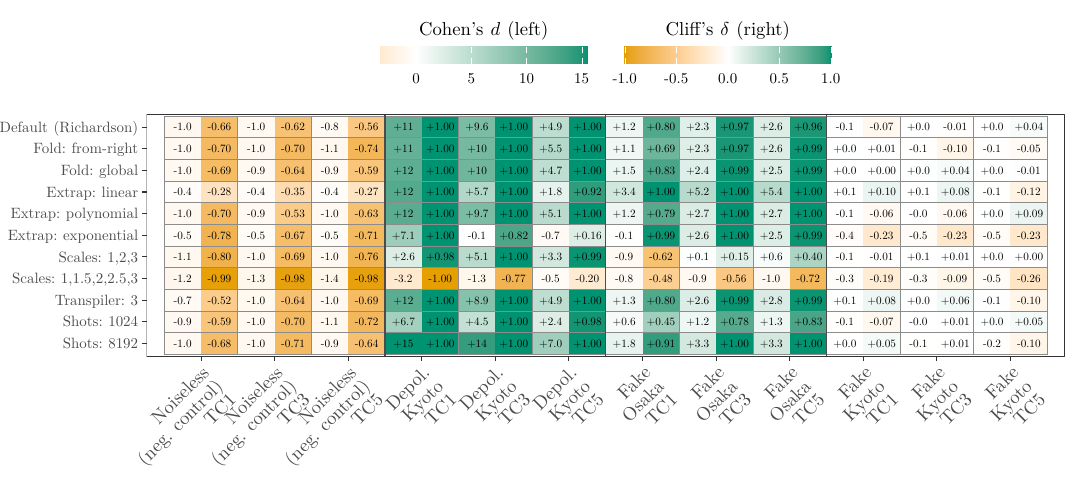}
  \caption{Parameter space heatmap for 4 backends $\times$ 3 Trotter depths (separated by vertical lines), one-at-a-time sweep from defaults.
  Each cell is split: the left half shows Cohen's $d$ (colour scale left), the right half shows Cliff's $\delta$ (colour scale right).
  Green = \gls{zne} significantly reduces error, yellow = \gls{zne} significantly increases error ($\alpha = 0.05$), white = not significant.}
  \label{fig:khan-heatmap}
\end{figure*}

For the missing parameters, we applied common default values: $4096$ measurement shots and transpilation level 1~\cite{qiskit2024}. Additionally, for \gls{zne}, folding-from-left strategy, Richardson extrapolation method, and scale factors \(\{1, 3, 5\}\) \cite{giurgica-tiron_digital_2020, majumdar_best_2023, cai_quantum_2023} are employed. From this baseline, we sweep values for alternatives and execute the configuration on calibration snapshots for IBM Kyoto and Osaka. Inspecting the Kyoto snapshot, we find ECR gate errors of all \(144\) qubits are near hundred percent, producing a noise-floor output. Given this is not the exact original experimental environment, we added a depolarising noise model matching the original work's error rate. We also added the reported QASM simulator for ideal results (negative control, as \gls{qem} cannot improve ideal results).

Ten configurations arise for a one-at-a-time sweep of the five parameters plus the baseline. We vary each axis independently while keeping the others at their default values and tests them against three Trotter depths (TC1, TC3 , TC5) specified in the original paper. In Summary $11$ configurations $\times$ $4$ backends $\times$ $3$ Trotter depths = $132$ configurations are tested in total.

\subsubsection{Statistical Analysis}
For each configuration, we run $n_\text{reps} = 200$ independent repetitions, then conduct paired $t$-tests on the per-repetition absolute error $|\epsilon|$ relative to the ideal expectation value, and compute Cohen's $d$ as an effect-size measure. We apply a significance threshold of $\alpha = 0.05$. 

\subsubsection{Artefact Classification}
Figure \ref{fig:khan-heatmap} shows the results of the parameter-space sampling. As expected, the ideal simulator shows a significant worsening with a median Cohen's $d$ of $-0.95$: this serves as a negative control, confirming that \gls{zne} cannot improve results that are already ideal, since any introduced noise amplification only degrades an already noiseless expectation value. In the last three columns we see neutral and mostly non-significant results for the fake IBM Kyoto (median Cohen's $d$ of $-0.03$), which is also expected given the fully faulty ECR gates~--~their errors drive the output to the totally mixed state, yielding near-zero expectation values regardless of the circuit. The other two backends show a more positive pattern. The depolarising simulation (Kyoto error rates from the paper) shows significant improvement in $29/33$ configurations with a median Cohen's $d$ of $+5.95$. Meanwhile the Osaka snapshot has a less positive improvement of median $+2.23$. As expected, the improvement is more pronounced for the depolarising simulation given its more idealised noise model. In contrast, the Osaka snapshot is closer to real hardware and therefore shows a more mixed pattern of results with other error patterns. The non-improving and smaller positive results are largely driven by the scale factors $\mathcal{S}$ (variance amplification, motivated in Section~\ref{sec:background}) and the exponential extrapolation method $\mathcal{E}$. Extrapolation methods assume different functional forms of noise decay and differ in their sensitivity to the noise model~\cite{endo_practical_2018,giurgica-tiron_digital_2020}: the exponential model assumes $E(\lambda) \propto (1-p)^{\lambda n}$, motivated by global depolarising noise~\cite{endo_practical_2018}. However, real hardware noise does not necessarily follow this model, as demonstrated in our case. $\mathcal{E}$ is therefore an active parameter whose noise-model sensitivity can change the conclusion about \gls{zne} effectiveness. Cliff's $\delta$ values (right half of each cell) closely mirror Cohen's $d$ patterns, confirming directional results are not an artefact of the shot-count inflation: Wherever $d$ is strongly positive, $\delta$ is near +1, and wherever $d$ is negative, $\delta$ is near -1.

\paragraph{Multiple-comparisons perspective}
Because repeated identical tests make false positives near-certain, we applied Bonferroni and Benjamini-Hochberg corrections to all paired $t$-test $p$-values. Of the 107 out of 132 uncorrected significant results (58 better, 49 worse), 103 survive the strict Bonferroni threshold (57 better, 46 worse) and 106 survive Benjamini-Hochberg. Only four results are dropped by Bonferroni, all corner cases near the noise floor. As both improvements \emph{and} degradations survive correction, experimental outcomes genuinely depend on implicitly assumed parameters, and the degradations on the noiseless and FakeKyoto backends are real rather than statistical artefacts.

\paragraph{Hardware calibration and effect size}
Khan~\etal report expectation values and variances (Table~III in~\cite{khan_error_2024}), but neither the statistical nature of these variances nor the number of repetitions per configuration is specified. We therefore cannot verify the reported improvements' significance or recover the paired differences for Cohen's $d$, and instead estimate $\hat{d}$ of the original paper by dividing $\Delta = E_\text{ZNE} - E_\text{raw}$ (from the reported values) by the simulated $\sigma_\text{improvement}$ as a proxy.

Since Khan~\etal used real IBM hardware, the IBM Osaka calibration snapshot is the closest available approximation for comparison. Our replication yields $d = +1.16$ at TC1~--~a moderate but statistically significant improvement. For the original paper~\cite{khan_error_2024}, $E_\text{raw} = 0.728$, $E_\text{ZNE} = 0.814$, $E_\text{ideal} = 0.828$; dividing the improvement $\Delta = 0.086$ by our $\sigma_\text{improvement} = 0.016$ gives $\hat{d} \approx +5.3$, more than four times our replication value. Under idealised depolarising noise (same circuit, same configuration), the same calculation yields $d = +11.3$. The original effect size therefore depends critically on the specific hardware calibration ($\mathcal{H}$) at the time of the experiment, which is no longer accessible. We therefore cannot distinguish whether $d\approx 1.16$, $d\approx 5.3$, or $d\approx 11.3$ is the better characterisation of the true effect, let alone assess its robustness across noise models~\cite{senapati_towards_2023,zheng_bayesian_2023}.

\begin{figure*}[htbp]
  \centering
  \includegraphics[width=\linewidth]{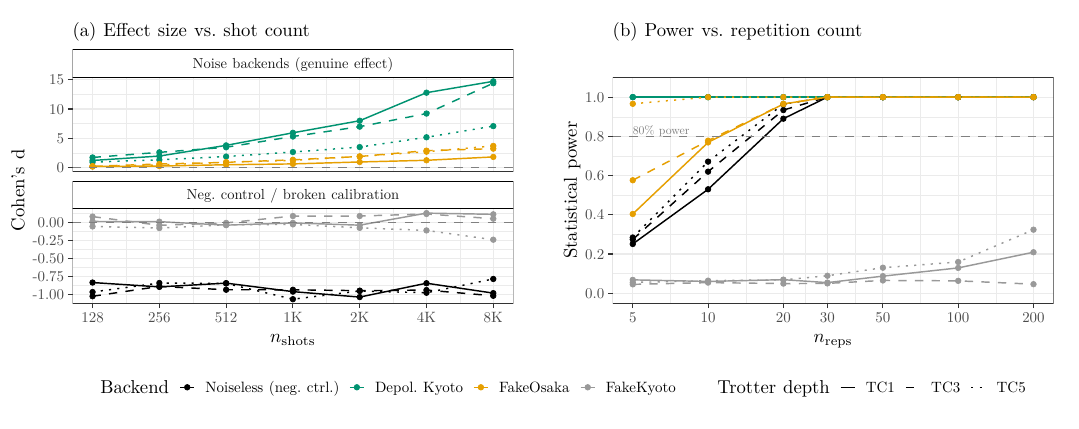}\vspace*{-2em}
  \caption{Sensitivity analysis.
  (a)~Cohen's $d$ vs.\ shot count; $d$ scales with $\sqrt{n_\text{shots}}$ for genuine effects.
  (b)~Statistical power vs.\ repetition count (1{,}000 bootstrap resamples); 80\% power requires $n_\text{reps} \geq 20$ for moderate effects.}\label{fig:khan-sensitivity}
\end{figure*}

To check if simulation-based findings transfer to quantum hardware, we ran the default configuration on the 156-qubit IBM Marrakesh processor~\cite{ibm_processor} for TC1 and TC3. This Heron machine's median two-qubit error rate of 0.241\% is roughly four times lower than IBM Kyoto's 0.947\%. We obtain $d = -0.75$ for TC1 but $d = +0.55$ for TC3: at TC1 the low error rate leaves the shallow circuit (six ECR gates) nearly ideal ($|\epsilon_\text{raw}| = 0.013$), so \gls{zne}'s variance amplification ($2.7\times$) outweighs any correction and pushes the mitigated value further from ideal, whereas TC3 (18~ECR gates) accumulates enough error for genuine improvement. This confirms hardware dependence and reveals a regime where low-error hardware makes \gls{zne} counterproductive for shallow circuits.

\paragraph{Shot-count variance and statistical power}
Figure~\ref{fig:khan-sensitivity}(a) confirms the expected $d \propto \sqrt{n_\text{shots}}$ scaling: on the depolarising model and FakeOsaka, $d$ grows monotonically with shot count, while FakeKyoto stays near zero (no genuine signal) and the ideal simulator is fixed at $d \approx -0.95$. A bootstrap power analysis (see Figure~\ref{fig:khan-sensitivity}(b)) shows that $n_\text{reps} \geq 20$ suffices for $80\%$ power at moderate effects, while large depolarising effects need as few as $n_\text{reps} = 5$; the near-zero FakeKyoto effect never reaches $80\%$ power even at $n_\text{reps} = 200$. Low shot counts ($\mathcal{Q}$) and few repetitions therefore risk both false negatives for genuine effects and an inability to confirm the absence of improvement where none exists.

\subsection{Case Study: Desdentado~\etal (2025)}
So far, we have assumed that multiple runs of an identical configuration yield the (essentially) identical result. Desdentado~\etal~\cite{desdentado_estimating_2025} provide a case where this assumption breaks: they observe a temporal confound whose structure is consistent with calibration drift, as
studied below in Section~\ref{sec:hardware_drift}. Their work proposes an algorithm to estimate the ideal shot count for a given quantum circuit to achieve the best possible result under \emph{shot noise}~--~the statistical sampling variance that decreases as $1/\sqrt{n_\text{shots}}$ when more measurement shots are taken. Hardware noise (\eg, gate errors, calibration drift,~\dots) is, in contrast, not reduced by increasing shot count.

\begin{figure}[htbp]
  \centering
  \includegraphics[width=\linewidth]{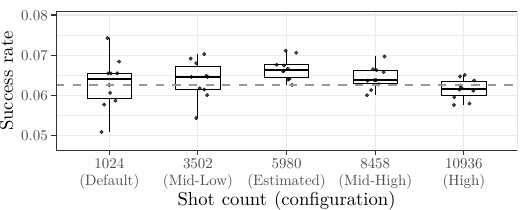}
  \caption{Shot estimation results from Desdentado~\etal~\cite{desdentado_estimating_2025} for a five qubit Grover circuit on the IBM Brisbane. The proposed shot estimation yields the best target state probability, while higher shot counts drift to the noise floor.}
  \label{fig:desdentado}
\end{figure}

The paper proposes an algorithm to estimate the ideal shot count and tests five configurations for a five-qubit Grover circuit on IBM Brisbane. If finds an improvement of $0.37\%$ for two, and $0.53\%$ for four target states. Their proposed shot estimation yields the best target state probability for all their circuits while lower and higher shot counts yield worse results. Uniquely, the paper provides a complete reproduction package that allows us to analyse exact hardware results from published data (see Figure~\ref{fig:desdentado}). Circuit depth and hardware error rate are dominant challenges: IBM Brisbane (Eagle r3) reports a median two-qubit gate error rate of $0.77\%$ error rate through the last IBM calibration snapshot. Alternate reported median ECR error rates ranging from $0.762\%$ \cite{robertson_simons_2025}, $0.79\%$ \cite{benito_comparative_2025} to $0.832\%$ \cite{abughanem_practical_2025}. The transpiled circuit uses 904 ECR gates, which leads to a total circuit fidelity of $\mathcal{F} = (1 - 0.00762)^{904} \approx 0.09\%$ with the most optimistic error rate.

The low circuit fidelity explains a near-random target state probability. Desdentado~\etal (see Figure~\ref{fig:desdentado}) report probability degradation after exceeding the estimated ideal shot count, with a non-monotonic pattern. We only show results for one circuit, but the pattern applies to all. This contradicts the shrinking variance of shot noise with increasing shot count, which should result in a monotonic improvement in target state probability as shot noise decreases. Job identifiers reveal that shot count groups were executed in sequence over a 20 minute time span, except for a gap of 13 minutes between parts of group 8458 and the entire group 10936. Because shot count is confounded with execution time and the group size is small ($n = 10$), inter-group differences may also reflect calibration drift. Particularly given the non-monotonic pattern (the last complete executed shot group, ideal shot estimation, coincides with lowest error), we cannot distinguish hardware drift, shot count or pure chance as cause. 

This illustrates how temporal drift can masquerade as parameter effect with un-randomised execution order. To isolate drift as an independent factor, we conduct a controlled longitudinal experiment measuring its impact on \gls{zne} effectiveness.

\emph{Cross-vendor fragility:} To test whether these findings transfer beyond IBM hardware, we executed the same circuit on the 54-qubit IQM Euro-Q-Exa~\cite{qexa} machine. At TC1 (6~CZ gates at $\lambda_1$), the circuit retains $27.7\%$ of the ideal expectation value ($\bar{E}(\lambda_1) = 0.290$ vs.\ $E_\text{ideal} = 0.980$). At $\lambda_3$ and $\lambda_5$, the expectation value is near the noise floor ($E(\lambda_3) = 0.024$) or negative ($E(\lambda_5) = -0.105$). TC3 (18~CZ) retains only $10.9\%$ of the ideal expectation value on Euro-Q-Exa, compared to $113\%$ on IBM Marrakesh (coherent over-rotation). The circuit depth at which \gls{zne} \enquote{works} is therefore a property of the circuit--hardware pair: cross-vendor studies are necessary for any generalisable claim. As the signal is near a fully mixed state at TC3 on the Euro-Q-Exa, TC1 is used for the longitudinal drift study in the next section.

\section{The Contingency of Time}
\label{sec:hardware_drift}
Even with fixed parameters, replications can depend on \emph{when} an experiment is run, given time-varying HW properties. We find only 28\% of papers address \emph{hardware drift} (C4). Given that superconducting processors exhibit calibration fluctuations for minutes to hours~\cite{hirasaki_detection_2023, senapati_towards_2023}, this raises concerns.

\subsection{Experimental Design}
A longitudinal experiment on the IQM Euro-Q-Exa system available at LRZ~\cite{qexa} tests the effect of temporal drift using the Khan~\etal~\cite{khan_error_2024} \gls{qtc} at depth TC1 (6 CZ at $\lambda_1$). We chose the smallest circuit, because the expectation value $\bar{E}(\lambda_1)$ of TC3 is already dominated by noise on this hardware. To minimise parameter confounds, we use default configurations from Section \ref{subsec:khan-params} and run an experiment every 30 minutes for different time periods.
This allows us to observe any temporal variability attributable to hardware drift. Four days with two 12 hour periods and one 48 hour weekend period with a total of $147$ time points with $n_\text{reps} = 30$ independent runs allow for observing short-term and long-term drift patterns.

We address three questions: Is drift reproducible across sessions (RQ1); do raw expectation values at $\lambda_1$ exhibit temporal autocorrelation inconsistent with the i.i.d.\ assumption of paired tests (RQ2); does drift cause per-time-point \gls{zne} effect size to vary substantially such that \emph{identical} experiments at different times yield different effectiveness \emph{conclusions} (RQ3).

\subsection{Results}
\label{sec:drift-results}
\begin{figure*}[htbp]
  \includegraphics{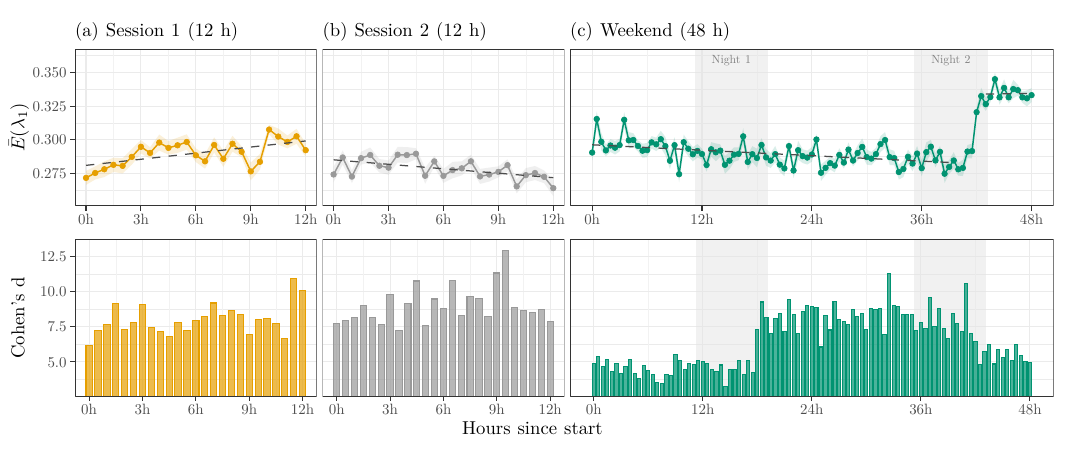}\vspace*{-1em}
  \caption{Longitudinal drift study on IQM Euro-Q-Exa (147~time points across 72~hours in three independent sessions).
  Top:~$\bar{E}(\lambda_1)$ averaged values with 95\%~CI vs.\ elapsed time; each session exhibits a qualitatively different drift pattern (step-change, gradual decline, overnight shift).
  Bottom:~per time point Cohen's~$d$ (\gls{zne} vs.\ raw): $d$ varies around $3\times$ across the weekend, this is a \emph{drift-induced effectiveness illusion}.}
  \label{fig:drift-combined}
\end{figure*}

Across all sessions, $\bar{E}(\lambda_1) \approx 0.28{-}0.30$, corresponding to approximately 29\% of the ideal expectation value $E_\text{ideal}$, $E(\lambda_3) \approx 0.02$ (near noise floor), and $E(\lambda_5)$ is consistently \emph{negative} (mean $-0.10$), which is inconsistent with depolarising noise decay and indicating coherent over-rotation past the zero-crossing. Figure \ref{fig:drift-combined} shows the time series across all three sessions alongside the associated \gls{zne} effect size per time point. Table~\ref{tab:drift-summary} summarises the drift severity metrics.

\begin{table}[htbp]
  \caption{Drift severity on Euro-Q-Exa.
  $\eta^2$: fraction of total variance between time points.
  $r_1$: lag-1 autocorrelation of  $\bar{E}(\lambda_1)$.
  $n_\text{eff}$: effective independent repetitions (nominal $n_{\text{reps}} = 30$).
  $d$~range: Cohen's $d$; all per time-point.}
  \label{tab:drift-summary}
  \centering
  \begin{tabular}{@{}lrrrrrl@{}}
    \toprule
    \textbf{Session} & \textbf{TPs} & $\boldsymbol{\eta^2}$ & $\boldsymbol{r_1}$ & $\boldsymbol{n_\text{eff}}$ & $\boldsymbol{d}$ \textbf{range} \\
    \midrule
    Day~1 (12~h)    & 25 & 0.35 & 0.55 & 3.5 & 6.1--10.9 \\
    Day~2 (12~h)    & 25 & 0.20 & 0.21 & 4.9 & 7.2--12.9 \\
    Weekend (48~h)  & 97 & 0.55 & 0.83 & 1.8 & 3.3--11.3 \\ 
    \bottomrule
  \end{tabular}
\end{table}

\begin{figure*}[htbp]
  \includegraphics{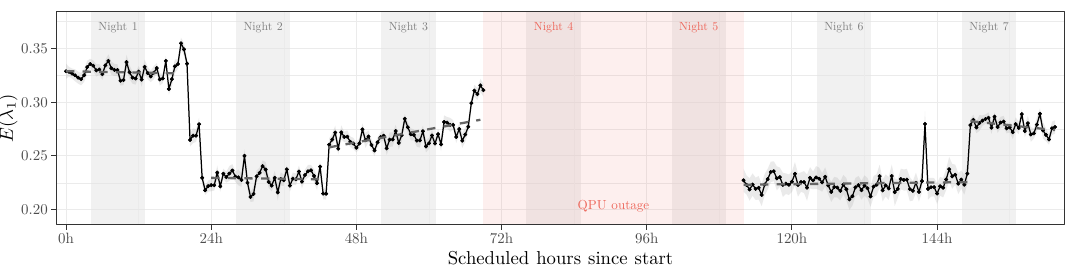}\vspace*{-0.5em}
  \caption{Seven-day longitudinal drift study on IQM Euro-Q-Exa (163~scheduled hours, 2026-04-08 to 2026-04-15).
  $\bar{E}(\lambda_1)$ per time point with 95\%~CI; dashed lines show piecewise-linear interpolation through the mean of each night-bounded daytime cluster (pre- and post-outage separately); grey bands mark local night (21:00--06:00~CEST) and the red gap a 43-hour QPU maintenance outage. $\bar{E}(\lambda_1)$ post-outage shows HW recalibration does not restore a stable baseline.}
  \label{fig:drift-week}
\end{figure*}

\paragraph{Three sessions, three drift patterns (RQ1)}
The top part of Figure~\ref{fig:drift-combined} reveals qualitatively different dynamics: Day 1 shows a slight upward drift with a discrete step-change at $t \approx 9.5$~h (that is, asymmetric across scale factors: $\lambda_1$ affected, $\lambda_{3,5}$ insensitive). Contrary to the first session, day 2 exhibits a gradual downward trend to the lowest measured expectation value. Session 3 was conducted over the weekend, and reveals what appears to be an overnight recalibration shift in the second night (between a Saturday and Sunday), which is absent in the first night. Between days 1 and 2, $E(\lambda_3)$ crosses zero ($+0.024 \to -0.011$), which the negative Richardson coefficients convert into a large change in the \gls{zne} estimate at around $t = 18$~h. Different patterns across sessions indicate that drift is not a stable, reproducible process, but a non-stationary phenomenon that can yield different outcomes for \gls{zne} effectiveness at different times. We extended this experiment to a full seven-day window (163~scheduled hours, Figure~\ref{fig:drift-week}): the trace shows qualitatively different overnight behaviour, and the baseline level after a 43-hour outage (red gap) differs visibly from before, confirming that drift persists over days and is not resolved by recalibration.

\paragraph{Drift is pervasive and severe (RQ2)}
In the 48-hour weekend study, more than half of the total variance in the raw signal is attributable to the time of measurement ($\eta^2 = 0.55$ (see Table \ref{tab:drift-summary})), and consecutive time points are strongly correlated ($r_1 = 0.83$). Each measurement carries information about the next, violating independence assumptions of standard paired tests. Additionally the autocorrelation is asymmetric: $r_1 = 0.17$ in the first 24~hours vs.\ $0.91$ in the second, coinciding with a visible upward shift during the second night ($t = 39{-}43$~h) which results in a higher $E(\lambda_1)$ up to $0.346$ compared to most of the other time points with expectation values below $0.300$. The 12-hour sessions confirm drift at lower severity ($\eta^2 = 0.20{-}0.35$, $r_1 = 0.21{-}0.55$).

\paragraph{The drift-induced effectiveness illusion (RQ3)}
Figure \ref{fig:drift-combined} (b) shows per-time-point Cohen's $d$ of \gls{zne} vs.\ raw across all sessions. The $d$ values vary substantially over time ($3.3$--$12.9$) (although inflated by the high measurement precision at $n_\text{shots} = 4096$). What matters is not effect size, but relative variation over time. For instance, in the weekend session, the experiment yields $d = 3.3$ at one time point and $d = 11.3$ twelve hours later: a $3.4\times$ difference in apparent \gls{zne} effectiveness. For comparison, switching from the Osaka calibration snapshot to a fundamentally different depolarising noise model, produces only a ratio of 2.7. \emph{Temporal drift on a single back-end can produce larger variation in apparent \gls{zne} effectiveness than changing the entire noise model}.

\begin{figure}[htbp]
  \includegraphics[width=\linewidth]{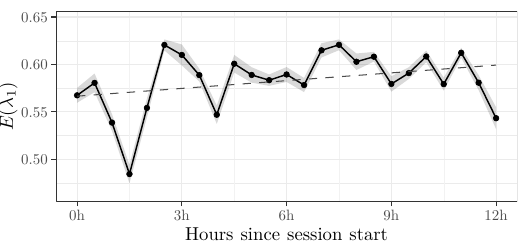}\vspace*{-0.5em}
  \caption{Twelve-hour drift session on IBM \texttt{ibm\_brussels} (TC3, 25~time points). $\bar{E}(\lambda_1)$ with 95\%~CI ribbon and linear trend (dashed); the dotted line marks the ideal value ($E_\text{ideal} = 0.846$). Despite a fixed hardware and parameter configuration, $\bar{E}(\lambda_1)$ drifts  across the session.}
  \label{fig:drift-ibm-brussels}
\end{figure}

In our experiment, the effect is significantly positive at every time point ($d > 3$ throughout), as expected: $\bar{E}(\lambda_1)$ is well above the noise floor and the high shot count inflates $d$ (Section~\ref{sec:bg-stats}). The illusion manifests not as a sign reversal, but as uncontrolled \emph{magnitude variation}: for moderate effects ($d \approx 1$--$2$) commonly reported in \gls{qem} hardware evaluations~\cite{kim_evidence_2023}, this temporal variation alone is sufficient to determine the binary conclusion of statistical significance. Additionally, the within-time-point \gls{icc} reduces the $n_\text{reps} = 30$ nominal repetitions to as few as $n_\text{eff} = 1.8$ effective independent observations (Table~\ref{tab:drift-summary}), following Kish's design-effect formula $n_\text{eff} = n / (1 + (n{-}1) \cdot \text{ICC})$~\cite{kish_survey_1965}. Note that this \gls{icc} is distinct from the between-time-point $r_1$: it is computed via one-way ANOVA \emph{within} each time point. A genuinely moderate improvement that appears highly significant at $n_\text{reps} = 30$ becomes non-significant once this effective sample size is accounted for. Two tests of the same \gls{zne} configuration~--~one in the morning, one the next day~--~can give contradictory conclusions. Neither would be wrong given measured results, but neither would paint an accurate picture. This limits generalisability of typical quantum (software) experiments that are often based on immediately consecutive hardware runs and limited repetition counts, especially given the financial commitment for such experiments.
IBM \texttt{ibm\_brussels} (TC3, 12-hour session, Figure~\ref{fig:drift-ibm-brussels}) shows $\bar{E}(\lambda_1)$ drifts by approximately $0.14$ within a single session, confirming that drift-induced effectiveness illusion is not specific to IQM.

\section{Discussion}
\label{sec:discussion}
\subsection{The compound nature of \gls{qem} artefacts}
Our analysis reveals that a favourable \gls{zne} outcome requires \emph{both} conditions to hold: the chosen parameter configuration must fall in the improving region of $\mathcal{P}$ (Section~\ref{sec:experiments_khan}), and the hardware calibration at the time of measurement must produce a stable, noise-discernible expectation value (Section~\ref{sec:hardware_drift}).

Critically, these two sources interact: a paper reporting a single configuration at one time point risks confounding both, as the outcome may reflect a fortunate parameter choice or a favourable calibration window. The Khan~\etal\ reproduction illustrates this compounding: even within the moderate Osaka noise model, $d$ ranges from $-0.75$ on IBM Marrakesh to $+11.3$ under idealised depolarising noise (Section~\ref{sec:experiments_khan}). Adding temporal drift introduces an additional $3.4\times$ variation in apparent \gls{zne} effectiveness within 48~hours on the same device.

This explains why few papers survive both challenges: a result must be robust against parameter choices \emph{and} stable under drift. The rarity of meeting all criteria suggests the literature overestimates \gls{zne} reliability not through method flaws, but because evaluation does not control for confounders.

\subsection{Scope and limitations}
We focus on \gls{zne} with Richardson extrapolation, the most widely used \gls{qem} method in our corpus. While the artefact sources~--~parameter sensitivity and temporal drift~--~are method-agnostic, our empirical evidence covers only \gls{zne}; structurally different techniques that remove manual scale-factor selection (\eg, probabilistic error amplification~\cite{kim_evidence_2023}) or that scale differently under drift, such as \gls{pec}~\cite{berg_probabilistic_2023} and Clifford data regression~\cite{czarnik_error_2021}, warrant separate validation before our quantitative claims are extended to them. Our one-at-a-time design does not capture interaction effects a full factorial design might reveal.

\emph{Internal validity:} Paired $t$-tests assume approximate normality of differences. While robust to moderate non-normality, smaller configurations may violate this assumption. We therefore computed the Wilcoxon signed-rank test for every configuration, where the tests agree on 129 of 132 outcomes, (all on the FakeKyoto backend at marginal $p$-values). Bonferroni and Benjamini--Hochberg corrections on 132 parameter-space configurations leave qualitative conclusions unchanged.

\emph{External validity:} Drift patterns may differ on other devices, architectures, or time scales. The Khan et al.\ reproduction uses noise-model snapshots rather than live hardware for the parameter sweep, which may not capture all device effects. 

\emph{Construct validity:} Our eight-criterion review framework is a pragmatic operationalisation of statistical rigour. Consequently, other framings could yield different compliance rates.

\subsection{Recommendations}
Based on our analysis, we propose a reporting checklist, ordered by implementation effort:
\begin{enumerate}[nosep,leftmargin=*]
  \item Document all \textbf{Active Parameters}, including calibration snapshot ($\mathcal{H}$), shot count and repetition count ($\mathcal{Q}$), transpilation seed, or method-specific hyper-parameters.
  \item Report \textbf{Inferential Statistics}: \emph{At least} pair claimed improvements with a hypothesis test and 
  effect-size measure, and treat an effect as practically meaningful only if it exceeds the variation induced by drift and parameter choice at the same device.
  \item Provide a \textbf{Reproduction Package} containing all code, data, transpiled circuits, calibration snapshots, etc. as explicit baseline for independent verification.
  \item Ensure \textbf{Result Robustness} by evaluating at least a small grid of configurations (\eg, multiple scale factors, back-ends,~\dots) to avoid misleading results.
  \item Quantify \textbf{Temporal Stability} by distributing HW experiments over in time, or randomise execution order to de-confound drift from parameter effects.
\end{enumerate}

\section{Conclusion}
\label{sec:conclusion}
Our systematic review of 81 \gls{qem} papers reveals predominantly descriptive reporting: only 25\% of the papers employ inferential statistics, and only 30\% address hardware drift. A two-stage analysis shows this methodological gap has substantial consequences: Prevailing practices yield different interpretations of the same technique when uncontrolled factors (\eg, parameter choice, execution time) vary.

First, we show implicitly assumed \gls{zne} parameters, including scale factors, extrapolation method, and hardware calibration, are \emph{active}: on two physics-based noise models, variations shift the conclusion from significant improvement to significant degradation in 8 of 66 configurations (12\%), while 55 (83\%) show significant improvement~--~depending solely on which parameters were implicitly assumed. We found that on IBM Marrakesh QPUs, \gls{zne} can even be counterproductive for shallow circuits whose unmitigated output is close to ideal. Second, our 72-hour longitudinal study on IQM Euro-Q-Exa shows temporal drift alone induces a 3.4-fold variation in apparent \gls{zne} effectiveness, exceeding the 2.7-fold variation observed when changing the entire noise model. The same drift reduces the $n_\text{reps} = 30$ nominal repetitions to as few as $n_\text{eff} = 1.8$ effective independent observations, substantially weakening the evidential basis of nominally repeated measurements.

Parameter sensitivity and temporal drift compound on real hardware. Their interaction challenges the validity of \gls{qem} benchmarks that do not include inferential testing, robustness analysis, and drift control. An apparent \gls{qem} improvement may reflect a favourable point in parameter space, a favourable calibration window, or both. This is not an issue of the methods, but of their \emph{use}: use-case-centric measurements are often carried out by domain specialists who reasonably rely on standard settings rather than low-level details. At its core this is a software challenge: better mechanisms, abstractions, and reference patterns would relieve \enquote{users} from such details. We hope our reproduction pipeline and reporting standards will support more robust \gls{qem} evaluation~--~and results with improved practical credibility and scientific soundness~--~as the field progresses towards practical quantum advantage.

\section*{Data availability}
Code, data, and plotting scripts are available in our \repro{} and permanent \zenodo{} that can build the paper including analysis results. HW calibration snapshots and logs allow for analysing parameters and drift without machine access.

\begin{small}\noindent\textbf{Acknowledgments}
The authors gratefully acknowledge the use of the quantum system Euro-Q-Exa, co-funded by the EuroHPC JU, BMFTR (grant 13N16690), and the Bavarian State Ministry of Science and the Arts, operated by the Leibniz Supercomputing Centre (LRZ) in Garching, Germany, for providing the computational resources for this work. We acknowledge partial support by the German Research Foundation, grant MA 9739/1-1, and the High-Tech Agenda of the Free State of Bavaria. We also acknowledge partial support by the European Union (Project Reference 101083427), the European Funds for Regional Development (EFRE) (Project Reference 20-3092.10-THD-105), by the European Regional Development Fund (ERDF) and by the Free State of Bavaria as part of the project AIM-SMEs (Grant No. 2506-014-3.2), co-funded by the European Union. 
\end{small}

\printbibliography

\end{document}